\documentclass[journal]{IEEEtran}
\usepackage{subfigure}
\usepackage{url}
\usepackage{pifont}
\usepackage[usenames]{color,colortbl}
 \definecolor{MyLightCyan}{rgb}{0,1.0,1.0}

\usepackage{amsmath, amsthm, amssymb, amsfonts}

\usepackage{graphicx}
\graphicspath{{image/}{pdf/}{fig/}{svg/}}
\DeclareGraphicsExtensions{.fig,.pdf,.png,.jpg}

\usepackage{figlatex}

\newcommand*\pair[2]{{<}{#1}{,}{#2}{>}}

\begin{document}
%
\title{Multi-Valued Routing Tracks for FPGAs in 28nm FDSOI Technology}

\author{
\IEEEauthorblockN{Sumanta Chaudhuri, Tarik Graba and Yves Mathieu} \\
\IEEEauthorblockA{Telecom ParisTech\\
Universit\'e Paris Saclay\\
46 Rue Barrault\\
Paris FRANCE 75634\\
Email: {firstname.lastname@telecom-paristech.fr}}
}
\maketitle

\begin{abstract}
In this paper we present quaternary and ternary routing tracks for FPGAs, and
their implementation in 28nm FDSOI technology.  We discuss the transistor level
design of multi-valued repeaters, multiplexers and translators, and specific features of
FDSOI technology which make it possible. Next we compare the multi-valued
routing architectures with equivalent single driver two-valued routing
architectures. We show that for long tracks, it is possible to achieve upto 3x
reduction in dynamic switching energy, upto 2x reduction in routing wire area
and $\sim$10\% reduction in area dedicated to routing resources. The
multi-valued tracks are slightly more susceptible to process variation.
We present a layout method for multivalued standard cells and determine the layout overhead.We conclude with 
various usage scenarios of these tracks.
\end{abstract}


%

\section{Introduction}

Multi-valued signalling has always been used in communication, mainly to
increase data rate. e.g  Techniques such as M-ary PAM (Pulse amplitude
modulation) are used in the Ethernet protocol. So far, on-chip communication
restricts itself to binary signalling, because the receiver and transmitter
complexity often outweighs the potential benefits of multi-valued signalling.
But as chips are growing bigger, and since the interconnect resistance does not
scale with technology~\cite{Miller}, interconnect (RC) delay plays an ever more
important role in chip design. Interconnect capacitance also increases from
generation to generation as lines are more closely spaced. To compensate for RC
delays, more buffers need to be inserted, giving rise to higher interconnect
power consumption.  For this reasons alternative interconnect technologies such
as optical, transmission lines etc. have been proposed.~\cite{Miller,tl}


In this article we focus on multi-valued communication within the chip to
alleviate some of the problems related to interconnects.  More specifically we
concentrate on routing tracks in FPGAs. Routing resources take up the most of
the area in an FPGA~\cite{Betz}, and also account for 60-70 \% of power
consumption~\cite{Shang}. Thus power efficiency of routing resources, and
decrease in routing congestion are always welcome.


Numerous researchers
have proposed ternary and quaternary logic designs over the
years~\cite{smith,hurst,dubrova}.  A majority of these designs use transistors
with several different $V_t$s (Threshold Voltage) which require different
amount of channel dopings and thus leading to complex processes. 
Some recent
implementation of MVL arithmetic circuits can be found in~\cite{tohoku}.
\cite{furber_ternary} proposes ternary signalling for asynchronous logic.
\cite{cunha,lazzari} proposes MVL lookup Table architectures. ~\cite{sige} proposes
ternary logic devices based on Si-Ge Diodes. One of the
successful commercialization of MVL is the strataflash from
Intel~\cite{strataflash1,strataflash2}.

In this article we discuss the use of multi-valued logic for on-chip
communications in FPGAs. To implement these circuits we use an interesting
feature of the contemporary FDSOI (Fully Depleted Silicon On Insulator)
technology, namely its capability to fine-tune the $V_t$ by varying the
Body-Bias.

\subsection{Organisation} The rest of the article is organised in the following
fashion. In section~\ref{sec:fdsoi} we provide a brief update on the FDSOI
technology, and discuss the back-biasing technique in detail.
In section~\ref{sec:mvl} we present the multi-valued logic systems and evaluate
the dynamic switching energies.  In section~\ref{sec:prim} we present the
various primitives such as repeaters, multiplexers and their operation.  In
section~\ref{sec:doe} we present the experimental architectures and
experimental methods to evaluate the performance of multi-valued tracks. In
section~\ref{sec:results} we present the experimental results and in
section~\ref{sec:var} we discuss the variability and reliability issues. In section 
~\ref{sec:layout} we present an outline of the method used to layout these circuits and we calculate the 
layout overhead. We conclude with various usage scenarios of these routing tracks.

\section{Brief Overview of FDSOI Technology}
\label{sec:fdsoi}
\begin{figure}
   \centering
   {\subfigure[FDSOI Transistor]{\includegraphics[height=0.1\textheight,width=0.3\textwidth]{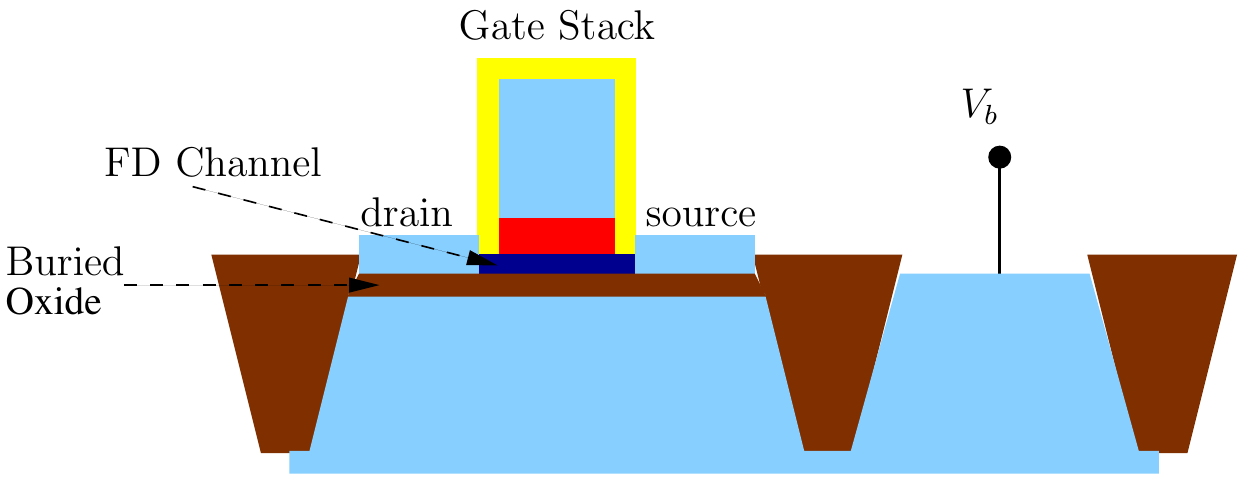}
   \label{fig:fdsoi1}}
   \subfigure[Back-biasing]{\includegraphics[width=0.15\textwidth]{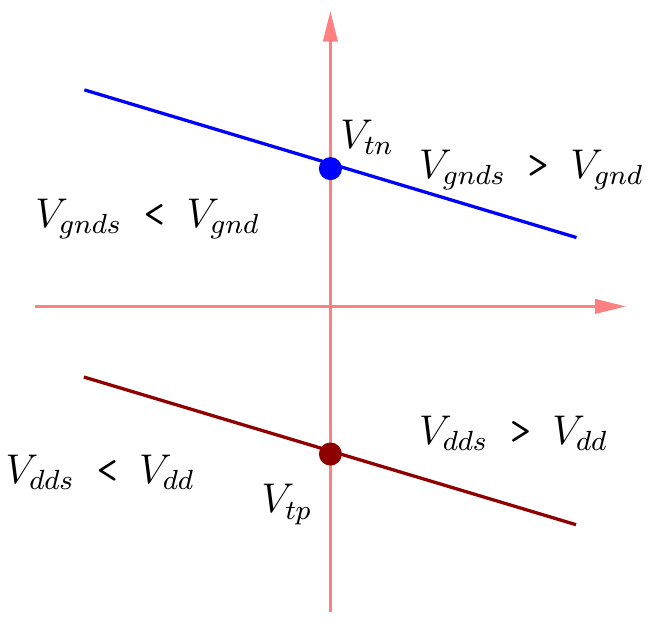}
   \label{fig:fdsoi2}}}
   \caption{Brief Overview of FDSOI Technology.}
   \label{fig:fdsoi}
\end{figure}

Figure~\ref{fig:fdsoi} depicts the structure of a FDSOI transistor. The main
difference with the bulk CMOS is the buried oxide (BOx) which insulates the
well from the channel. The silicon layer/channel is fully depleted, that is, it
does not contain any active charge carriers. Some of the potential benefits of
this structure are

\begin{itemize}
\item
Improved Junction capacitance.
Lower parasitic capacitance, i.e lower source-drain capacitance thanks to
dielectric isolation.
\item
Better Electro-static control of the channel, which results in a near ideal
sub-threshold slope of 60mv/decade~\cite{fdsoi_ieee_proc}, and reduced DIBL
(Drain induced Barrier Lowering)

\item
Improved $V_t$ variation as the channel is not doped. One of the major causes
of $V_t$ variation is RDF (Random Dopant Fluctuation). Thus variability
coefficient for transistors with same size is 2-3 times less for
FDSOI~\cite{cheng09}

\item
The transistor  is controlled through two independent gates. The $V_t$ can be
modulated by applying back-bias to the back plane (BP).

\end{itemize}

In FDSOI the $V_t$ can be controlled through three different methods
\begin{itemize}
\item
By doping the substrate below the Buried Oxide (BOx), also known as back plane
(BP). When the BP is of opposite polarity to source/drain the $V_t$ increases.
It is know as the RVT (Regular $V_t$ ) flavor.  When the BP is of same polarity
as source/drain, we get the LVT (Low $V_t$) flavor and it has a lower $V_t$.

\item
The $V_t$ can be controlled by applying a voltage to the backplane, however to
avoid formation of forward biased diodes in the substrate there are some
limitations.  Figure:~\ref{fig:fdsoi2}  shows the back biasing characteristics.
The $V_t$ is lowered during Forward Body Bias (FBB) corresponding to $V_{dds} <
V_{dd}$ for PMOS and $V_{gnds} >  V_{gnd}$ for NMOS. Similarly $V_t$ can be
augmented using Reverse Body Bias (RBB).
\\
\begin{tabular}{|c|c|c|}
\hline
RVT & RBB upto -3V & FBB upto +300mv\\ \hline
LVT & FBB upto +3V & RBB upto -300mv\\ \hline
\end{tabular}
\\
\item
It is also possible to increase $V_t$ by increasing the gate length. This is
known as poly-biasing.
\end{itemize}

In this article we will use a combination of the above three methods to
implement multi-valued logic.  For more details about UTBB-FDSOI please refer
to~\cite{stfdsoi}.

\section{Multi-Valued Logic Systems (MVL)}
\label{sec:mvl}
While ternary systems are the optimum~\cite{trit}, there are many advantages of implementing
arithmetic circuits in a multi valued logic systems~\cite{smith}. In this
article we consider mainly quaternary systems, but the circuits presented 
can be easily adapted to ternary systems.

On top of arithmetic advantages, there are some physical advantages related to
these systems, particularly when used to communicate over a long distance.
They are namely dynamic switch energy and routing congestion.
\begin{itemize}
\item
The dynamic switching energy in the context of on-chip communication for
quaternary logic is detailed in table~\ref{tab:transitions}. We can
see that average energy /transition for quaternary is $0.27\times Cvdd^2$. Similarly it 
can be shown that average energy /transition for ternary systems is
$0.33 \times Cvdd^2$.
\item
For quaternary logic a reduction of 50\% in routing wire area is achievable.
For ternary logic systems upto 33\% reduction in routing wire area can be
achieved. This calculation is based on the fact that 2 ternary wires can
transmit as much information as 3 binary wires.
\end{itemize}

\begin{table}[h]
\begin{center}
\caption{Energy for different transitions in a 4-valued signal, IN0(=$0\times vdd$),IN1($=\frac{1}{3}\times VDD$), IN2(=$\frac{2}{3}\times VDD$), IN3(=$VDD$).}
\label{tab:transitions}
\begin{tabular}{|c|c|}
\hline
\multicolumn{2}{|c|}{4-Valued transitions}  \\ \hline
Transitions & Energy \\ \hline
$t_{00}$,$t_{11}$,$t_{22}$,$t_{33}$ & 0 \\ \hline
$t_{01},t_{12},t_{23},t_{32},t_{21},t_{10}$ & \tiny{{$C\times \frac{1}{9} \times vdd^{2}$}} \\ \hline
$t_{02},t_{13},t_{20},t_{31}$ & \tiny{$C\times \frac{4}{9}\times vdd^{2}$} \\ \hline
$t_{03},t_{30}$  & \tiny{$C\times vdd^{2}$} \\ \hline
Av. Energy/Tran  & \tiny{$0.27 \times C vdd^{2}$} \\ \hline
\end{tabular}
\\
\vspace{2em}
\begin{tabular}{|c|c|}
\hline
\multicolumn{2}{|c|}{2-Valued transitions}  \\ \hline
transitions & Energy \\ \hline
$t_{00}$,$t_{11}$ & 0 \\ \hline
$t_{01},t_{10}$ & \tiny{{$C\times vdd^{2}$}} \\ \hline
Av. Energy/Tran  & \tiny{$0.5 \times C vdd^{2}$} \\ \hline
\end{tabular}
\end{center}
\end{table}
%
\section{Primitives for Multi-Valued Routing}
\label{sec:prim}
\subsection{Down-Literal Converters}
\label{sec:dlc}
\begin{figure*}[!t]
\centering
\includegraphics[width=0.7\textwidth]{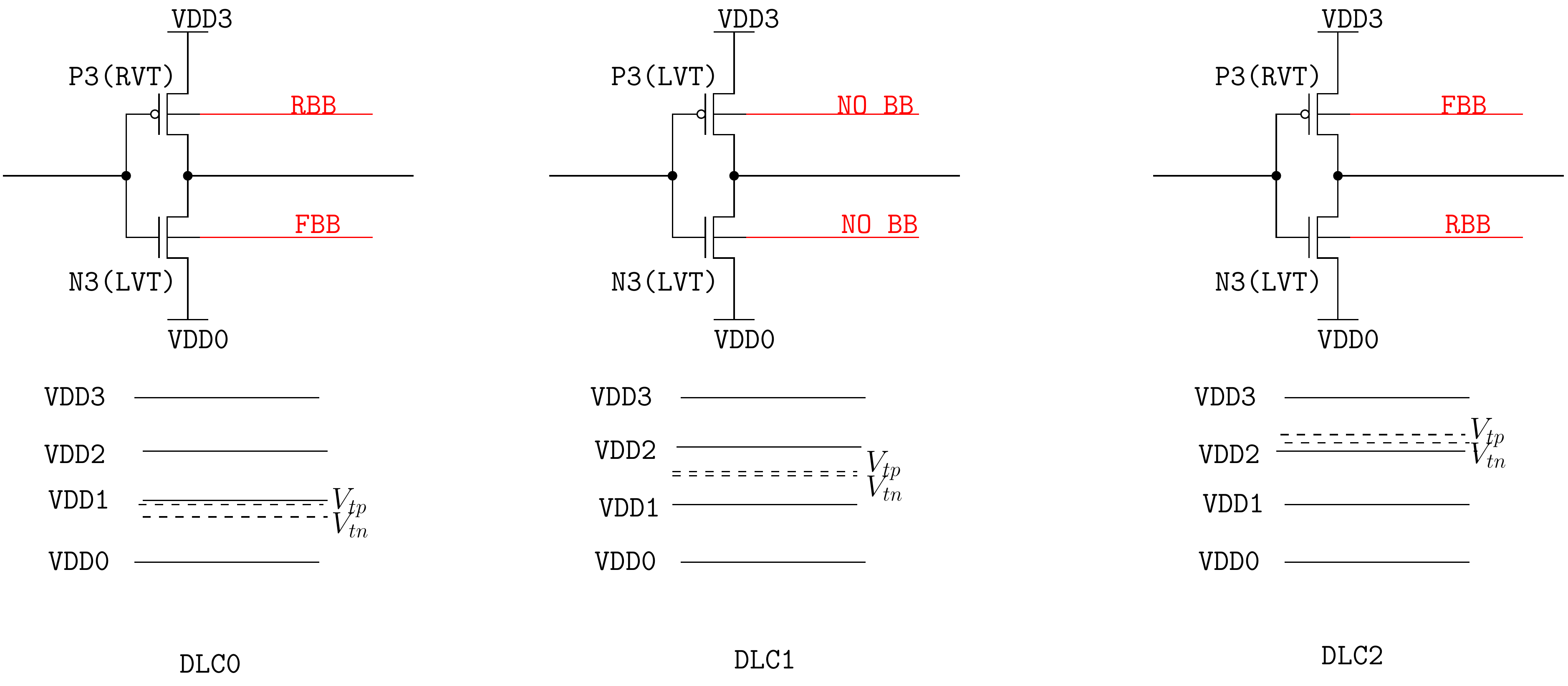}
\caption{Down Literal Converters for 4-valued Logic}
\label{fig_dlcs}
\end{figure*}

\begin{table}
\begin{center}
\label{tab:dlc}
\caption{Down-Literal Converters, 0(=$0\times VDD$),1($=\frac{1}{3}\times VDD$), 2(=$\frac{2}{3}\times VDD$), 3(=$VDD$).}
\begin{tabular}{|c|c|c|c|}
\hline
\multicolumn{4}{|c|}{Down-Literal Converters}  \\ \hline
input & DLC0 & DLC1 & DLC2 \\ \hline
0 & 3 & 3 & 3 \\ \hline
1 & 0 & 3 & 3 \\ \hline
2 & 0 & 0 & 3 \\ \hline
3 & 0 & 0 & 0 \\ \hline
\end{tabular}
\end{center}
\end{table}
Down literal converters (DLCs) are the basic primitives of MVL implementation
with binary CMOS logic~\cite{smith}.  They help in categorizing  the multiple
valued signals in different bins. Table~\ref{fig_dlcs} describes the three down
literal converters associated with Quaternary logic. DLC0 distinguishes between
logic level $<0>$ and $<1,2,3>$. DLC1 distinguishes between $<0,1>$ and $<2,3>$
and so on.  In figure~\ref{fig_dlcs} we can see implementation of DLCs in FDSOI
technology with the use of back biasing. For DLC0 the threshold voltage
$V_{tp}$ is increased by using a RVT PMOS transistor and Reverse Back-biasing,
and $V_{tn}$ is engineered to be between logic level 0 and 1. In the same
fashion other DLCs are implemented. It is also possible to use poly-biasing 
(see sec. ~\ref{sec:fdsoi}) to increase the reverse bias.

\subsection{Multi-Valued Multiplexers}
\label{sec:mux}
\begin{figure}[!h]
   \centering
   {
   \subfigure[Multi-Valued Multiplexers with binary select inputs. VDDBOOST=1.1V, VDD3=0.9v]{\includegraphics[width=0.3\textwidth]{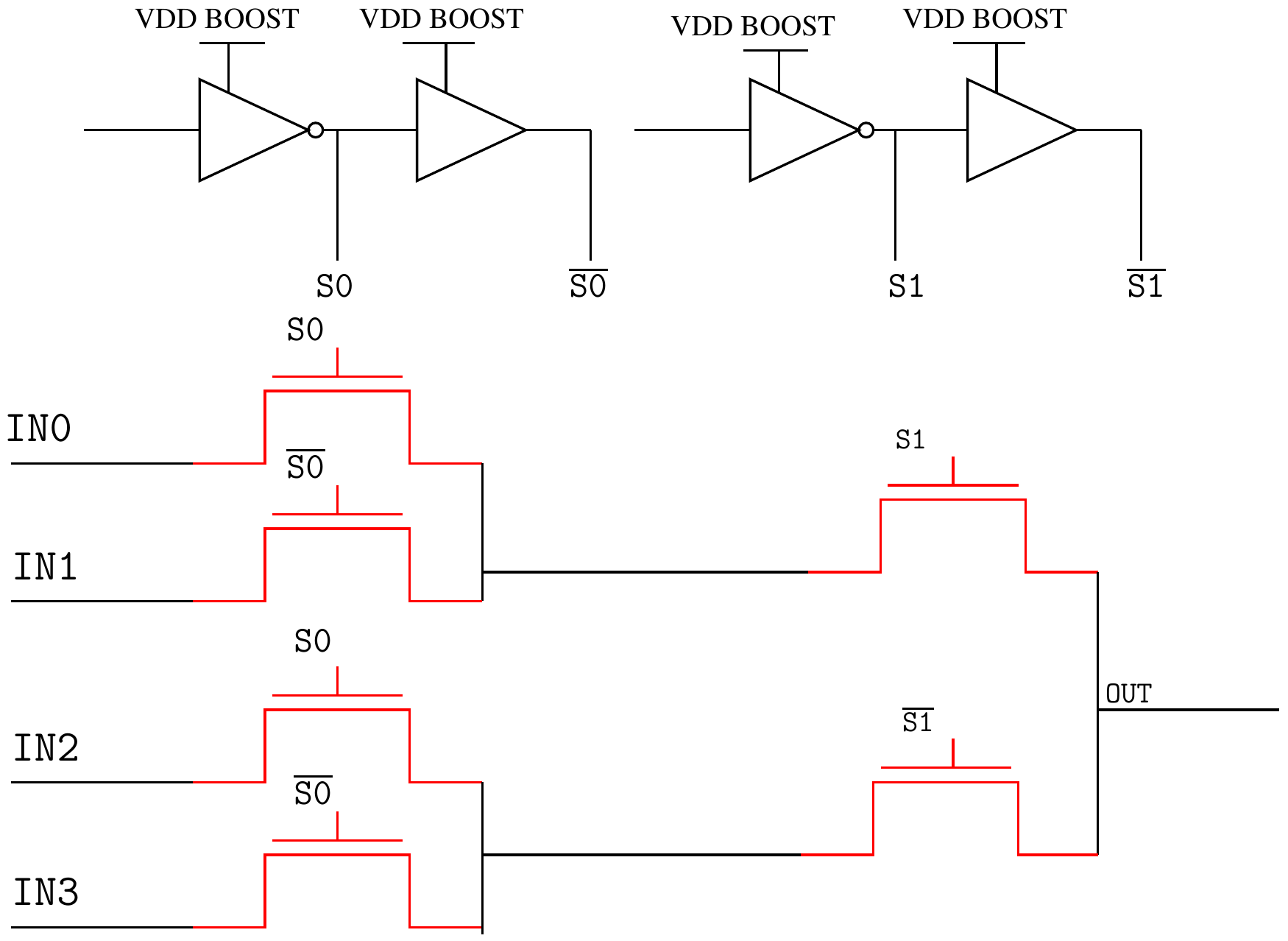}
   \label{fig:4vl_2vl_mux}}}
   \caption{Multiple-Valued Multiplexer Circuits for Routing.}
   \label{fig:mux}
\end{figure}

The multiplexer described in figure~\ref{fig:mux} assumes that
inputs and outputs are multi-valued, whereas the select input is binary. In
this case we assume that the configuration memory points are actually 6T binary
SRAM cells.

The use of boosted gate voltages
is commonplace to avoid any signal degradation for logic level '1'. The same
gate-boosted transistor can also pass all the levels in a multiple valued
logic. It is also possible to use transmission gates instead, but we consider
pass transistors as they  have lower transistor count.

\subsection{Multi-Valued Repeater Circuits}
\label{sec:repeater}

\begin{figure}[]
   \centering
   \includegraphics[width=0.45\textwidth]{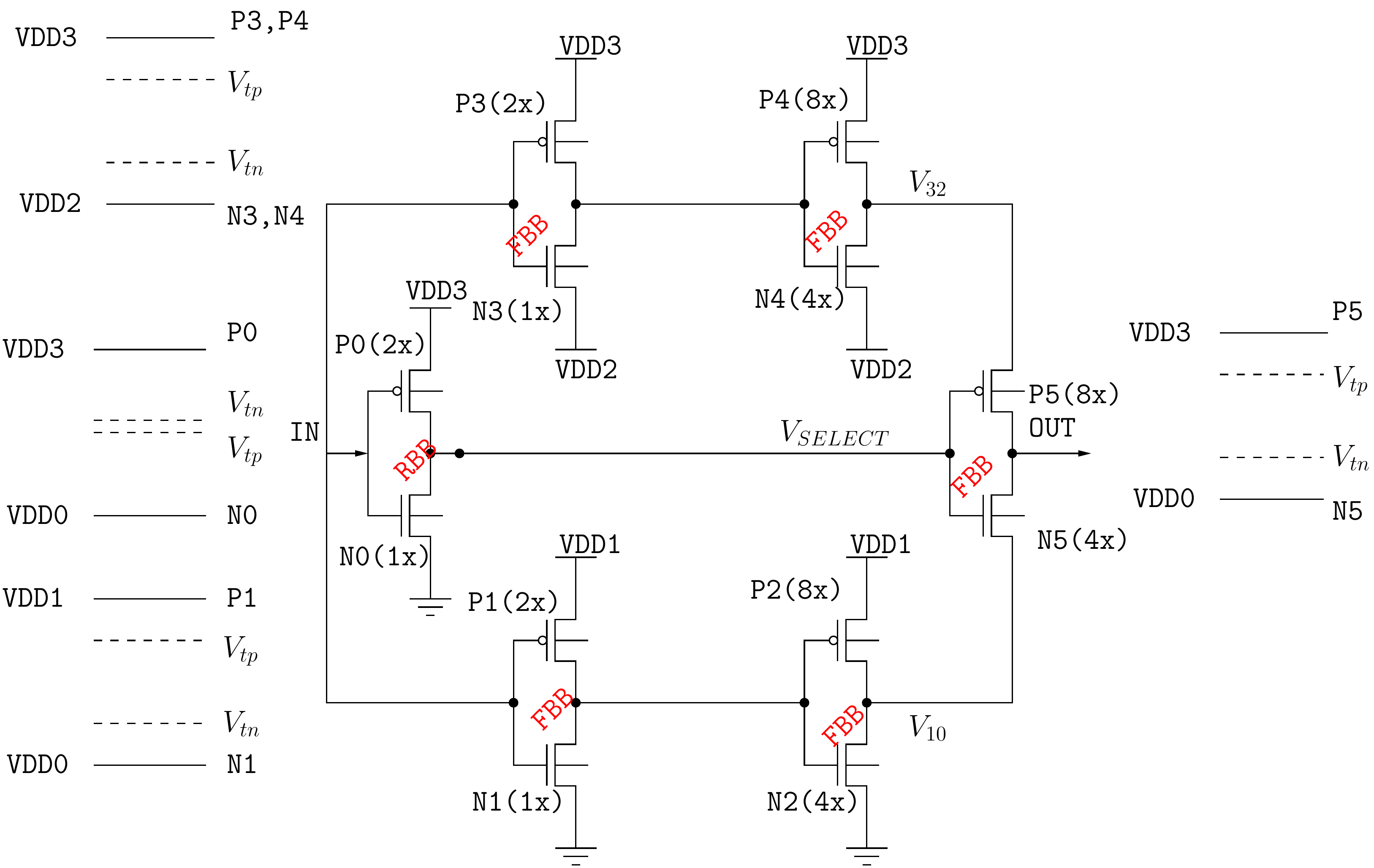}
   \caption{Quaternary Repeater Circuits for Routing,VDD3=0.9V, VDD2=0.6V, VDD1=0.3V.}
   \label{fig:repeater4}
\end{figure}

\begin{figure}[b]
   \centering
   \includegraphics[height=0.3\textheight,width=0.5\textwidth]{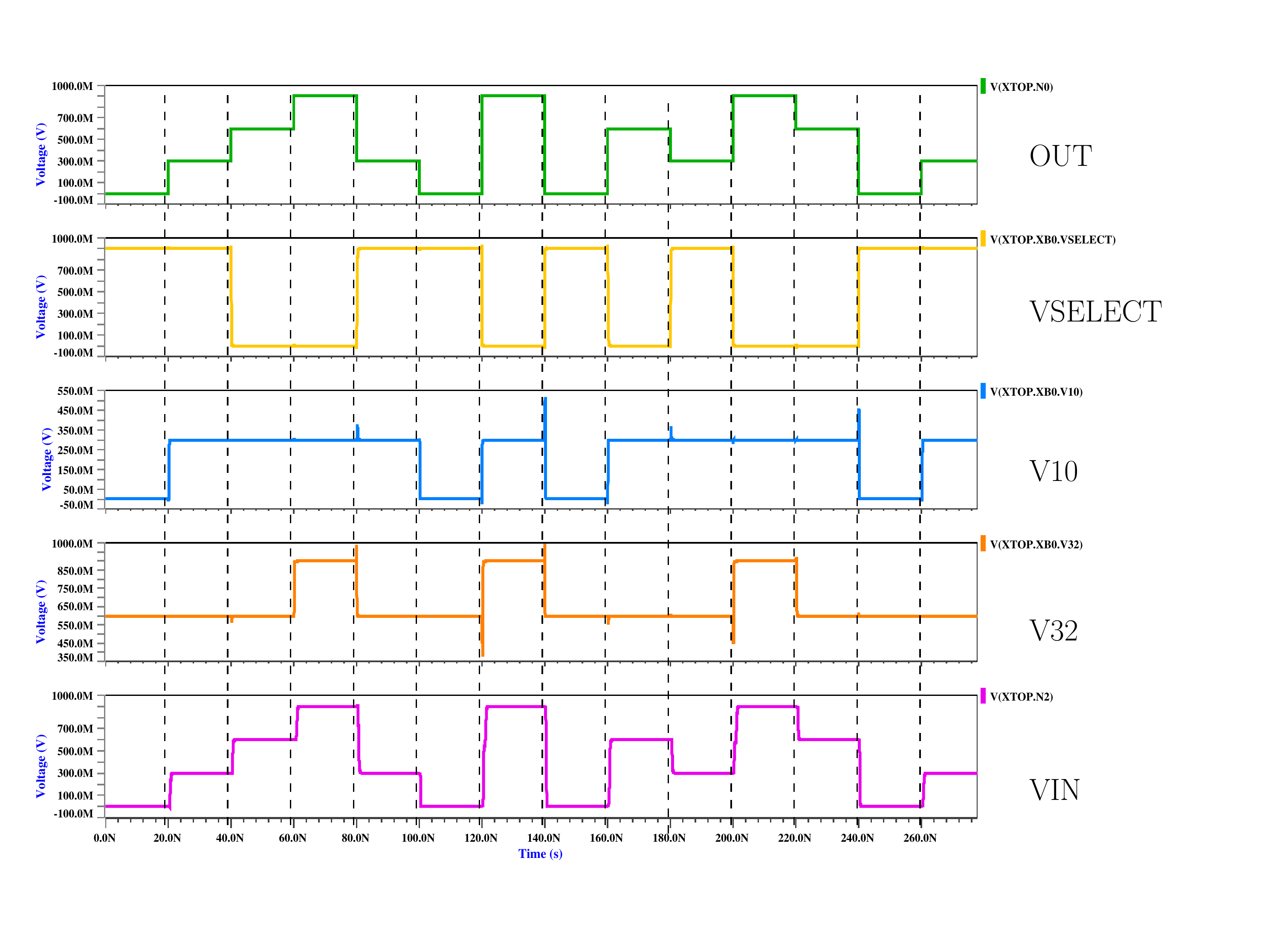}
   \caption{SPICE (ELDO) Simulation results for the Quaternary Repeater. Waveforms for $V_{32}$, $V_{10}$ \& $V_{SELECT}$ illustrates the operation of the repeater. (corresp. figure~\ref{fig:repeater4})}
   \label{fig:repeater_plot2}
\end{figure}

\subsubsection{Operation}

\newcolumntype{K}[1]{>{\columncolor[gray]{0.8}}c}

{\tiny
\begin{table}[t]
   \begin{center}
      \caption{Operation of 4-valued Repeater, IN0(=$0\times VDD$),IN1($=\frac{1}{3}\times VDD$), IN2(=$\frac{2}{3}\times VDD$), IN3(=$VDD$).}
      \label{tab:repeater}
      \begin{tabular}{|p{0.8 em}|K{0.8 em}|K{0.8 em}|p{0.8 em}|p{0.8 em}|K{0.8 em}|K{0.8 em}|p{0.8 em}|p{0.8 em}|K{0.8 em}|K{0.8 em}|p{0.8 em}|p{0.8 em}|}
         \hline
         \tiny{IN} &      \tiny{N0} &      \tiny{P0} &      \tiny{N1} &      \tiny{P1} &      \tiny{N2} &      \tiny{P2} &      \tiny{N3} &      \tiny{P3} &      \tiny{N4} &      \tiny{P4} &      \tiny{N5} &      \tiny{P5} \\ \hline
         \tiny{0} &      \tiny{OFF} &       \tiny{ON} &      \tiny{OFF} &       \tiny{ON} &       \tiny{ON} &      \tiny{OFF} &      \tiny{OFF} &       \tiny{ON} &       \tiny{ON} &      \tiny{OFF} &       \tiny{ON} &      \tiny{OFF} \\ \hline
         \tiny{1} &      \tiny{OFF} &       \tiny{ON} &       \tiny{ON} &      \tiny{OFF} &      \tiny{OFF} &       \tiny{ON} &      \tiny{OFF} &       \tiny{ON} &       \tiny{ON} &      \tiny{OFF} &       \tiny{ON} &      \tiny{OFF} \\ \hline
         \tiny{2}  &       \tiny{ON} &      \tiny{OFF} &       \tiny{ON} &      \tiny{OFF} &      \tiny{OFF} &       \tiny{ON} &      \tiny{OFF} &       \tiny{ON} &       \tiny{ON} &      \tiny{OFF} &      \tiny{OFF} &       \tiny{ON} \\ \hline
         \tiny{3}  &       \tiny{ON} &      \tiny{OFF} &       \tiny{ON} &      \tiny{OFF} &      \tiny{OFF} &       \tiny{ON} &       \tiny{ON} &      \tiny{OFF} &      \tiny{OFF} &       \tiny{ON} &      \tiny{OFF} &       \tiny{ON} \\ \hline
      \end{tabular}
   \end{center}
\end{table}
}

Figure~\ref{fig:repeater4} depicts the circuit diagram for  the quaternary repeater circuit consisting of 12 transistors.
The engineered \emph{Vth}s of each transistor pair are indicated in the diagram.
The transistors $\pair{P0}{N0}$ selects among the level $\pair{0}{1}$ and $\pair{2}{3}$.
When the input belongs to $\pair{0}{1}$ the transistor $N5$ is on and net $V_{10}$ is connected to the output.
When the input belongs to $\pair{2}{3}$ the transistor $P5$ is on and net $V_{32}$ is connected to the output.
The transistors $\pair{P1}{N1}$, $\pair{P2}{N2}$ act as a buffer for the $\pair{0}{1}$ levels and  the transistors $\pair{P3}{N3}$, $\pair{P4}{N4}$ act as a buffer for the $\pair{2}{3}$ levels.

The states of each transistor for various levels at the input are described in table~\ref{tab:repeater}.
Detailed operation of the quaternary repeater can be understood from the waveforms of its internal nodes presented in figure~\ref{fig:repeater_plot2}.

The ratio of average dynamic switching energy of such a repeater w.r.t 2 binary tracks can expressed as:

{\small
\begin{equation}
   \frac{E(switching)_4}{E(switching)_2} = \frac{(C_{wire}+3\times C_{L})\times 0.27 \times Vdd^2}{2\times (C_{wire}+C_{L})\times 0.5 \times Vdd^2}
\end{equation}
}

Where $C_L$ is the input capacitance of the transistor pairs at the input of the repeater.
The input capacitance of a quaternary repeater is three times that of a binary buffer.
For this reason with short track lengths where the energy consumption and delay are much higher.
As the track length grows  $C_{wire} \gg 3\times C_{L}$ the ratio of dynamic energy approaches $0.27$.


This gain in energy comes with an associated increase in delay.

Here we present a simplified analysis of the repeater delay.
In our simplified model, the resistance and the capacitance of an interconnect line are replaced by a effective capacitance ($Ceff_{wire}$) driven by the gate~\cite{Qian_ceff}.
$Ceff_{wire}$ is a function of the interconnect R\&C.
We assume charging of this effective capacitance  by a current source (i.e MOS).
Although this is an  approximate model it is adequate to explain the design trade-offs.

In general, with above assumptions, given the same transistor dimensions, and the same input slew rate, the delay (high to low) is proportional to $\frac{C_L \times Vdd}{(Vdd-Vtn)^\eta}$ where $\eta$ is between 1 and 2 for newer technologies due to velocity saturation~\cite{rabaey}, and $Vtn$ is the NMOS threshold voltage.

The ratio of delay can be thus expressed as

{\small
\begin{equation}
   \frac{\tau_{4}}{\tau_{2}} = \frac{(Ceff_{wire}+3C_{L})\times(Vdd/3)}{(Vdd/3-Vtn)^\eta} \times \frac{(Vdd-Vtn)^\eta}{(Ceff_{wire}+C_{L})\times(Vdd)}
\end{equation}
}

Note that we have considered the worst case delay in the quaternary case, i.e transitions where the voltage swing is lowest $Vdd/3\rightarrow 0$.

For long wires where $Ceff_{wire} \gg 3C_L$ we can simplify the above to

{\small
\begin{equation}
\frac{\tau_{4}}{\tau_{2}} = \frac{3^{\eta-1}}{(Vdd-3\times Vtn)^\eta} \times (Vdd-Vtn)^\eta
\end{equation}
}

In the quaternary repeater we have forward biased the driving stage transistors lowering their Vth.
So we need to slightly modify the above equation, where $Vtn'$ denotes the modified $Vth$ for NMOS transistor.

{\small
\begin{equation}
   \frac{\tau_{4}}{\tau_{2}} = 3^{\eta-1} \times \frac{(Vdd-Vtn)^\eta}{(Vdd-3\times Vtn')^\eta}
\end{equation}
}

\begin{figure}[h]
   \centering
   \includegraphics[width=0.5\textwidth,angle=0]{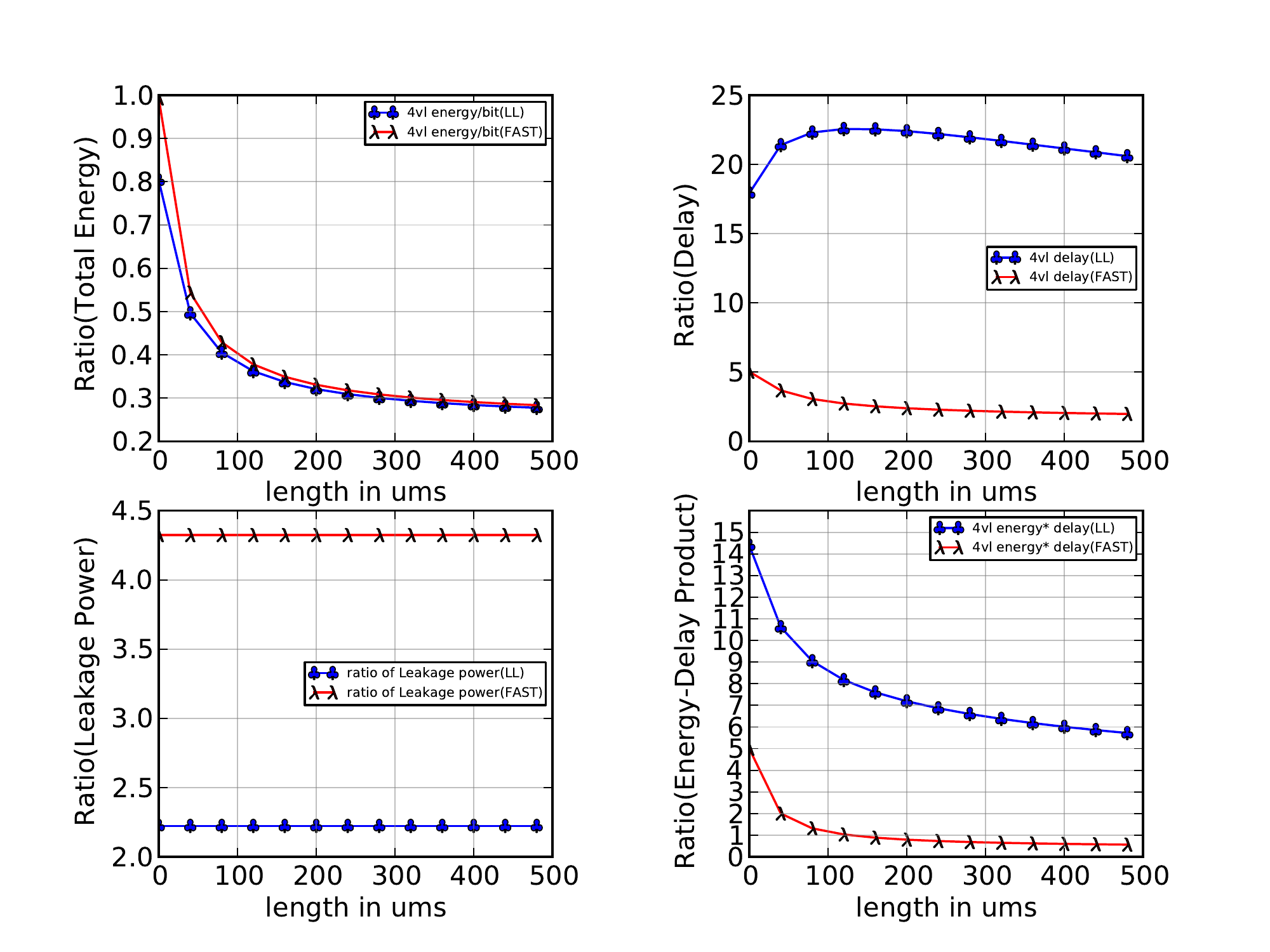}
   \caption{Energy-delay plots with varying routing track length, in $\mu$ms. The plotted value is the ratio with equivalent binary architectures.}
   \label{fig:results}
\end{figure}

In case we forward bias the driving transistors such that $Vtn'=\frac{1}{3} \times Vtn$ we can expect around 80\% increase in delay.
However there will be a considerable increase in leakage power ($\sim$ 4x) as can be seen from fig.~\ref{fig:results} (assuming $\eta$ to be equal to $1.5$).

Figure~\ref{fig:results} plots the energy and delay comparison of the quaternary and binary repeaters.
We have optimized the 4-valued repeater circuits by varying the back-biasing with various different goals, namely FAST is optimized for delay, and LL is optimized for low leakage.

The energy delay plots of these repeaters with varying routing track length is shown in figure~\ref{fig:results}.
They plot the ratio of energy and delay for multi-valued repeaters w.r.t equivalent configurations with binary buffers.
For quaternary, one 4-valued wire is compared with 2 binary wires, transmitting the same information.
In the binary case we use a tapered 2 stage binary buffer with the 1x transistors for input stage, and 4x transistor for output stage.
The tracks are terminated with standard four inverter loads.
We simulated  a single repeater driving a track of varying length, with a test vector where all transitions (as shown in table.~\ref{tab:transitions}) are equally represented.

The test-vectors use a cycle time of 10ns (100MHz) for FAST and 70ns(15 MHZ) for LL. This information is used to separate the leakage power from the dynamic power.
The plotted values are ratios of equivalent quantities w.r.t the value measured in the base architecture.

We have also plotted the energy delay product with varying length of tracks, and we see indeed at longer track lengths multi-valued signalling gets interesting.
Please note, that these calculations are done at 100MHz (15MHz for LL) , so the effect of leakage is mitigated.
For low operating frequencies(~KHz) the energy delay product will be largely dominated by leakage power.

\subsection{Multi-Valued to Binary Translator Circuits}
\label{sec:xlat}
\begin{figure}[]
   \centering
   {\subfigure[Quaternary to Binary Translator]{\includegraphics[width=0.35\textwidth]{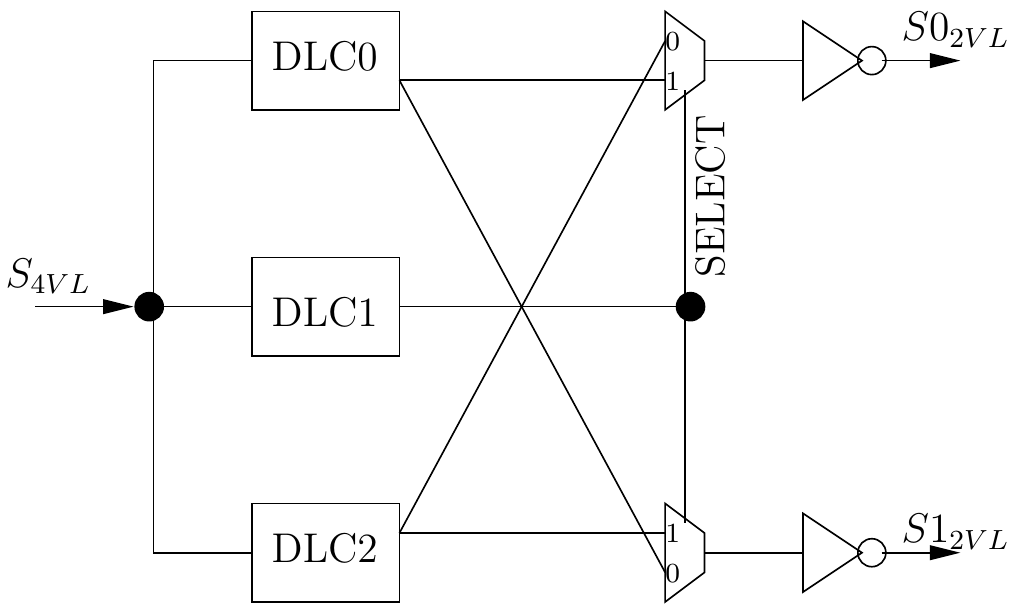}
   \label{fig:xlat42}}
   \subfigure[Binary to Quaternary Translator]{\includegraphics[width=0.4\textwidth]{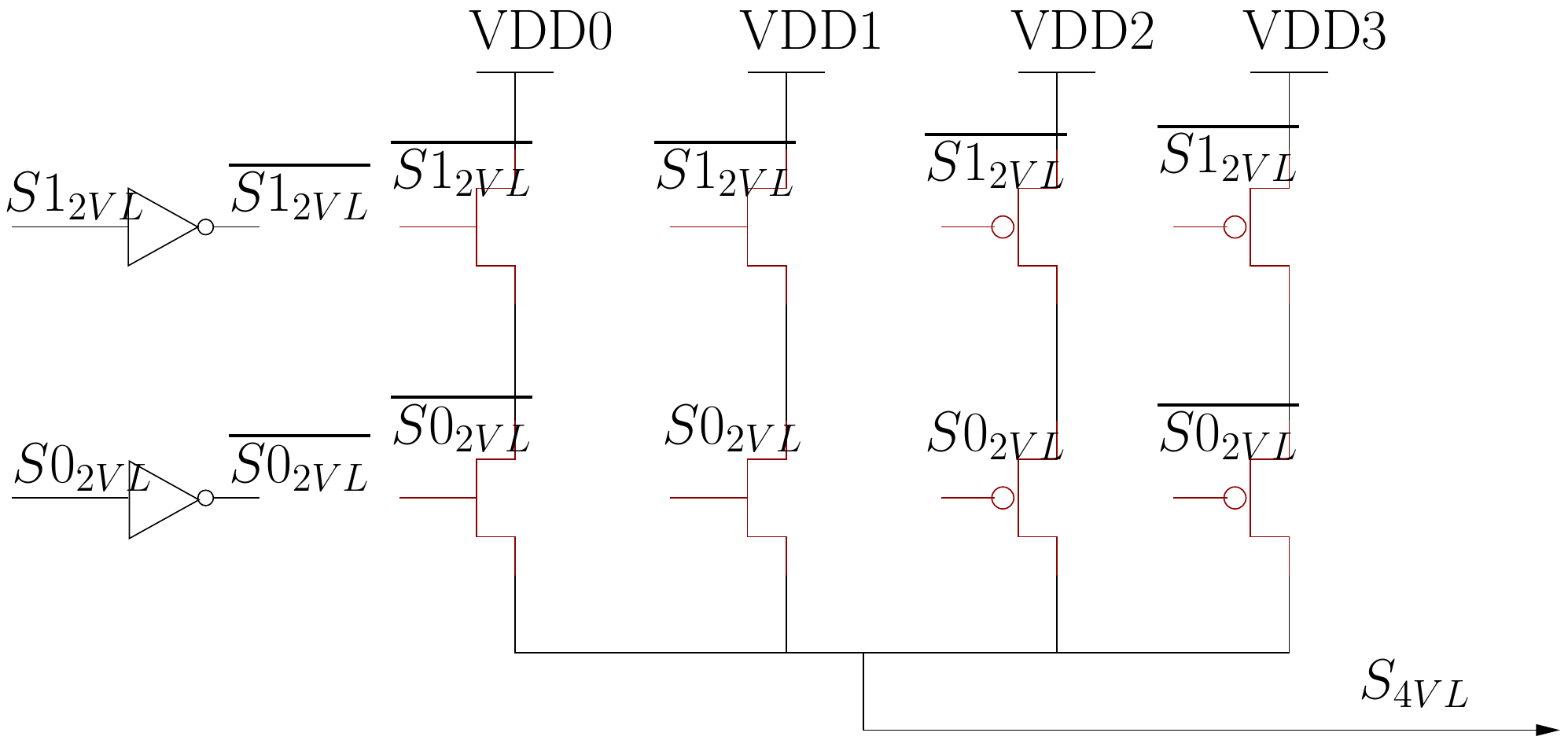}
   \label{fig:xlat24}}}
   \caption{Quaternary to Binary Translators and Vice-Versa.}
   \label{fig:xlat}
\end{figure}

The multi-valued routing tracks can be used in the context of a end-to-end
multi-valued FPGA, or it can also be used in a binary FPGA to implement bus-based
tracks~\cite{rose_bus}. In the later scenario we will need binary to quaternary translators
and their counterparts.
In fig.~\ref{fig:xlat} we present the binary/Quaternary translators. The design can be extended
to include ternary translators as well.  

In Fig.~\ref{fig:xlat42} we use DLCs to convert a quaternary 
signal to two binary signals. The output of the DLC1 is used to control the multiplexer and the signal levels 
are detailed in table ~\ref{tab:xlat}.

\tiny
\begin{table}[h]
\begin{center}
\caption{Operation of 4-to-2 translator, 0(=$0\times VDD$),1($=\frac{1}{3}\times VDD$), 2(=$\frac{2}{3}\times VDD$), 3(=$VDD$).}
\label{tab:xlat}
\begin{tabular}{|p{2.6 em}|p{2.6 em}|p{2.6 em}|p{2.6 em}|p{2.6 em}|p{2.6 em}|p{2.6 em}|}
\hline
\tiny{$S_{4VL}$} &      \tiny{DLC0} &      \tiny{DLC1} &      \tiny{DLC2} &      \tiny{SELECT} &      \tiny{S0} &      \tiny{S1} \\ \hline
\tiny{VDD0} &      \tiny{VDD3} &       \tiny{VDD3} &      \tiny{VDD3} &       \tiny{1(VDD3)} &       \tiny{1(VDD3)} &      \tiny{1(VDD3)}     \\ \hline
\tiny{VDD1} &      \tiny{VDD0} &       \tiny{VDD3} &       \tiny{VDD3} &      \tiny{1(VDD3)} &      \tiny{0(VDD0)} &       \tiny{1(VDD3)}      \\ \hline
\tiny{VDD2}  &       \tiny{VDD0} &      \tiny{VDD0} &       \tiny{VDD3} &      \tiny{0(VDD0)} &      \tiny{1(VDD3)} &       \tiny{0(VDD0)}      \\ \hline
\tiny{VDD3}  &       \tiny{VDD0} &      \tiny{VDD0} &       \tiny{VDD0} &      \tiny{0(VDD0)} &      \tiny{0(VDD0)} &       \tiny{0(VDD0)}      \\ \hline
\end{tabular}
\end{center}
\end{table}
\normalsize

Fig.~\ref{fig:xlat} depicts the binary-to-quaternary translator. 
The quaternary levels (i.e \emph{VDD0},\emph{VDD1},\emph{VDD2},\emph{VDD3}) are routed
to the output depending on the input binary values ($S1,S0$). We can see from the diagram that only one of the paths
can be activated at one time.

\section{Design of Experiment}
\label{sec:doe}

In this section we present the experiments to compare power consumption and
delay of Multi-Valued FPGA architectures and the base architecture, which is a
binary mesh FPGA with single-driver tracks~\cite{lemieux}. Single driver tracks
are most common in modern FPGAs due to their reduced capacitative loads and
better area-delay performance.

\subsection{Base and Experimental Architectures}
\label{subsec:arch}
\begin{figure}[]
   \centering
   {\subfigure[Base Architecture, Switch Box and Connection Box with Single-Driver BUS2 tracks]{\includegraphics[width=0.30\textwidth]{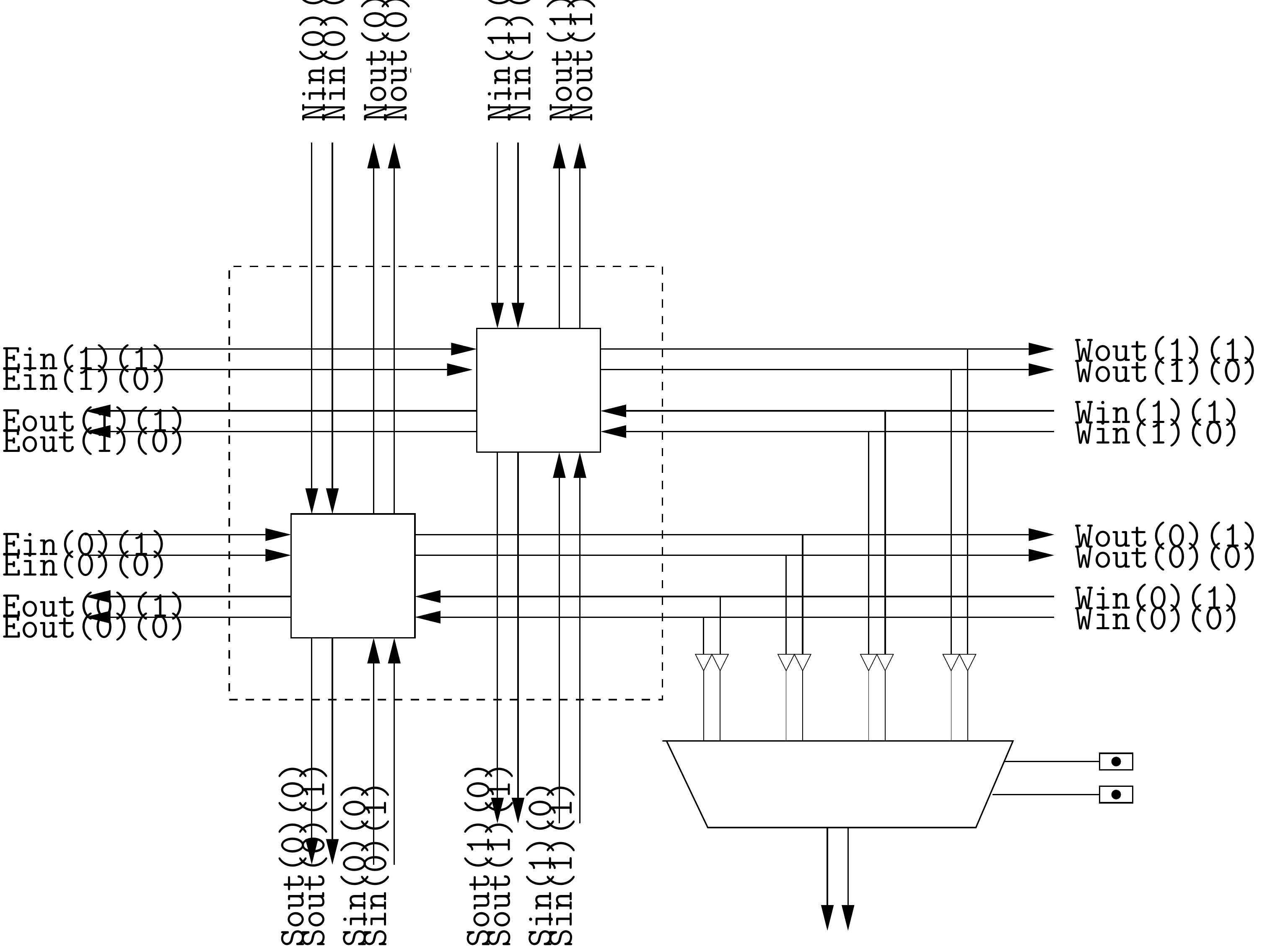}
   \label{fig:base_switchbox}}
   \subfigure[Base Architecture: Detailed view of a Switch Point for Bus-2 tracks]{\includegraphics[width=0.30\textwidth,angle=0]{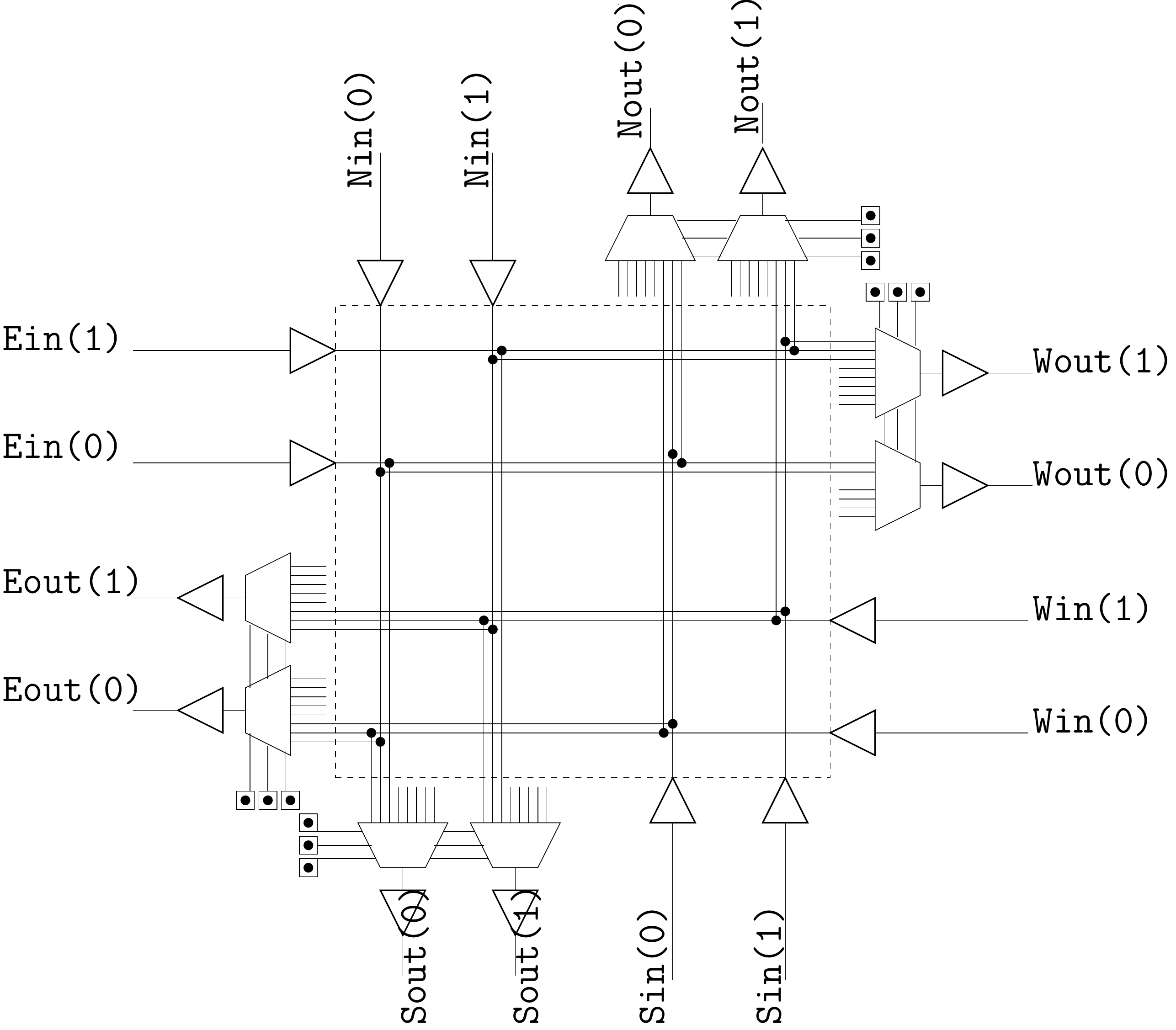}
   \label{fig:base_switchpoint}}
   \subfigure[Multi-valued Architecture: Switch box and Connection box with single driver quaternary tracks]{\includegraphics[width=0.30\textwidth,angle=0]{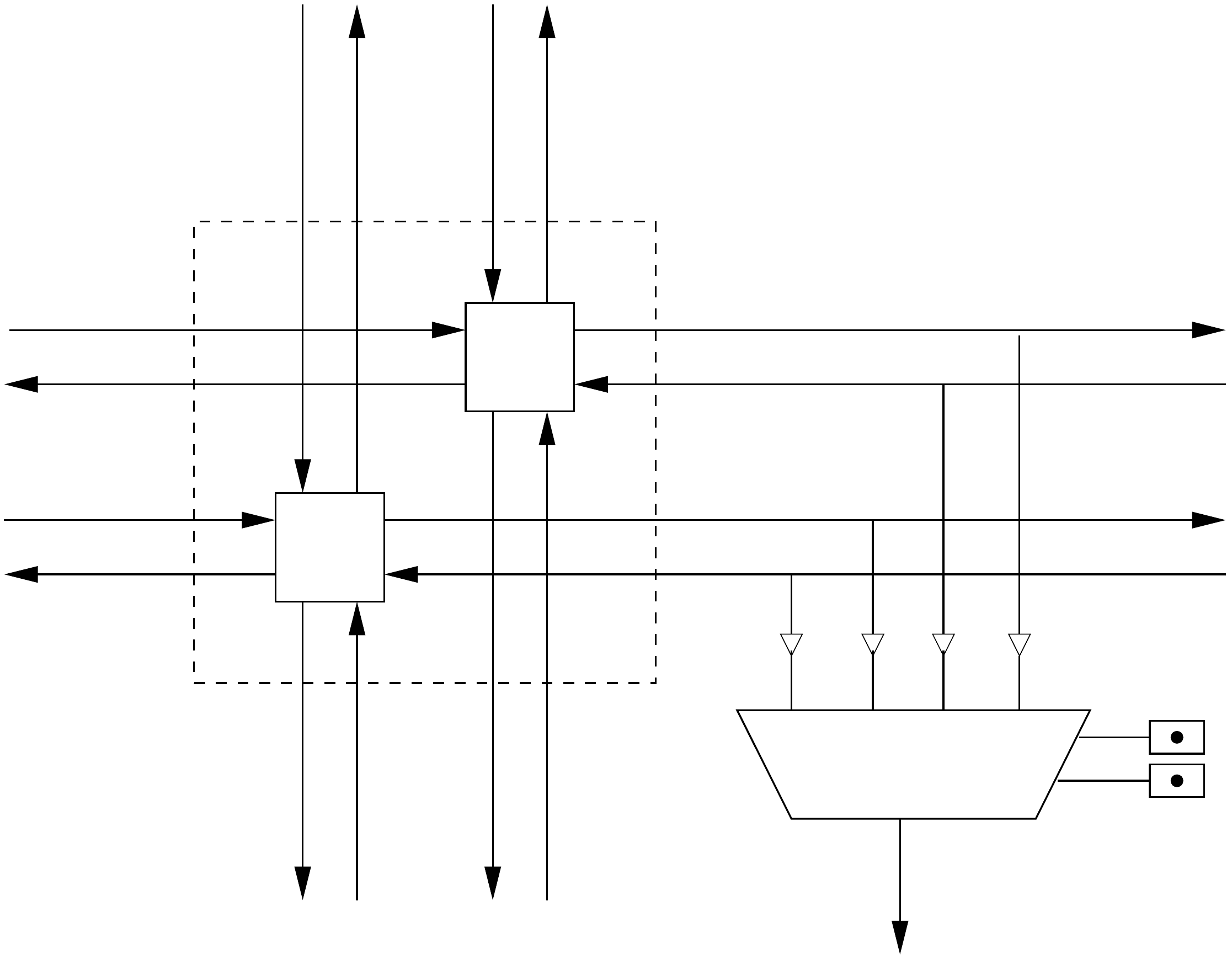}
   \label{fig:mvl_switchbox}}
   \subfigure[Multi-Valued Architecture: Detailed view of a Switch Point for quaternary tracks]{\includegraphics[width=0.30\textwidth,angle=0]{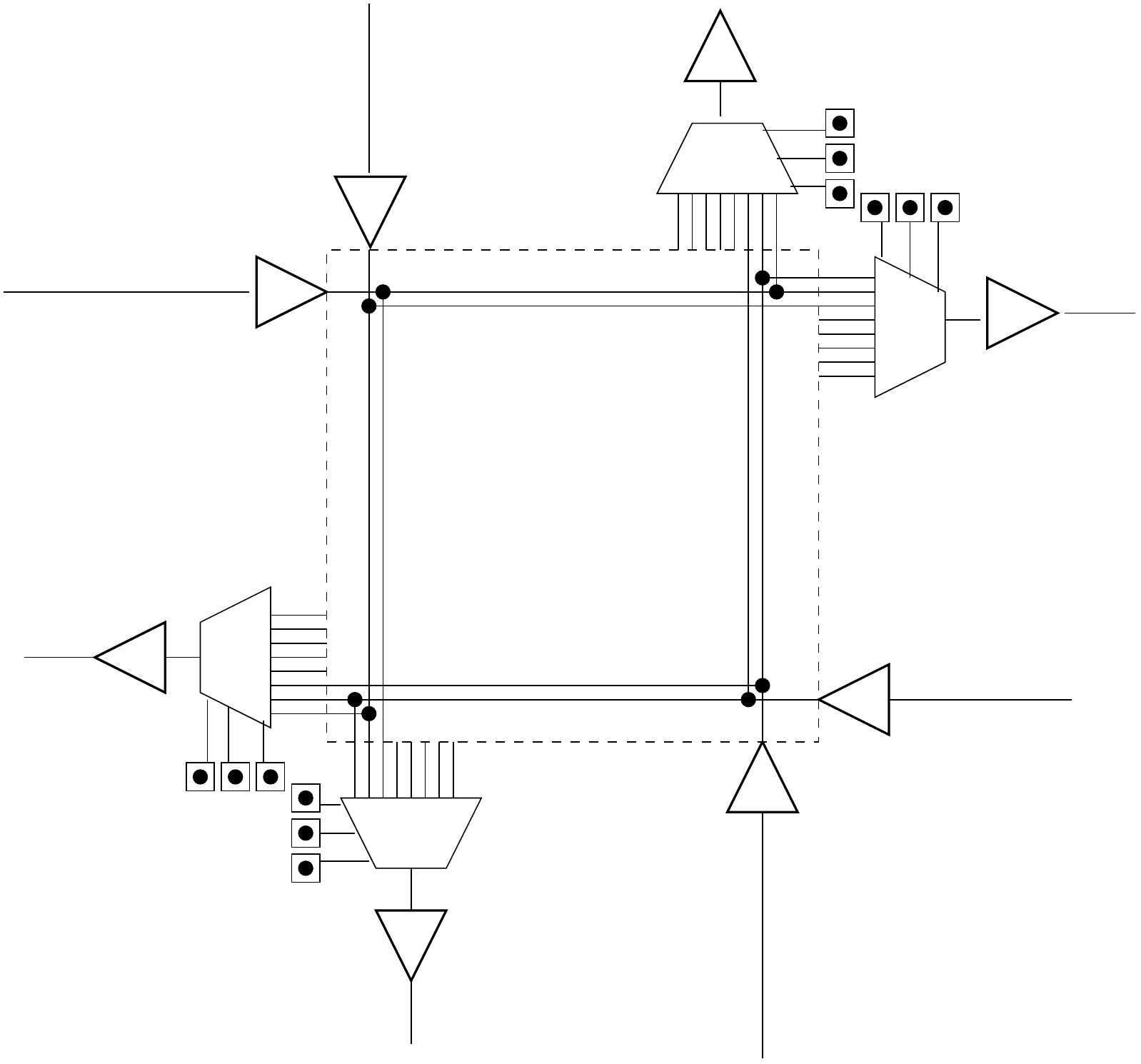}
   \label{fig:mvl_switchpoint}}}
   \caption{Base and MVL architectures used in the experiments.}
   \label{fig:arch}
\end{figure}

In our base architecture, as depicted in figures~\ref{fig:base_switchbox}  a
single-driver bus of width 2 is used to carry 2 bits of information equivalent
to a quaternary signal. This makes both architectures functionally equivalent.
As in the single driver architecture,  the output of logic blocks connect
directly to the switchbox mux~\cite{lemieux}.  A detailed view of the switch
point in base architecture is described in figure~\ref{fig:base_switchpoint}.
For routing a Bus-2 track the switch-points require double the number of muxes
and buffers.

Figure~\ref{fig:mvl_switchbox} describes the equivalent architecture in
quaternary, and the detailed view of the switch point is described in
figure~\ref{fig:mvl_switchpoint}. This is similar to binary FPGAs, but each
track is carrying 2 bits of information and quaternary buffers are used. For
muxes we assume the use of multi-valued muxes with binary select inputs (see
section~\ref{sec:mux}), as the configuration memory points are still binary.

For tracks we have a unit track length equal to the width of a basic FPGA tile
(CLB+Switchbox). The track capacitance and resistance are extracted with the
Cadence tool QRC.

For our experiment we assumed the following architecture:
\begin{itemize}
\item
No. of LUTs+FFs per CLB=4
\item
No of Inputs/CLB=16, No of Outputs/CLB=4.
\item
No. of Tracks (W)=64
\item
Input Connection Box Flexibility=0.25. i.e each CLB input
connects to 16 tracks among the 64 tracks.
\end{itemize}
Table~\ref{tab:tcount} details the transistor counts for each basic block and
compares them with traditional architectures.
\begin{table*}
\begin{center}
\caption{Transistor count in Basic blocks.}
\label{tab:tcount}
\begin{tabular}{|c|c|c|c|}
\hline
\multicolumn{2}{|c|}{\textbf{Binary Bus-2 Routing Tracks}} & \multicolumn{2}{|c|}{\textbf{Quaternary Routing Tracks}}  \\ \hline
Muxes(Bus-2) with N select inputs & $4*N+ 2*(2^{N+1}-2)$ & Mux with N binary select inputs & $4*N+ (2^{N+1}-2)$  \\ \hline
8:1 Muxes(Bus-2) with 3 select inputs & 40T & 8:1 Mux with 3 select inputs & 26T  \\ \hline
16:1 Mux(Bus-2) with 4 select inputs & 76T & 16:1 Mux with 4 select inputs & 46T  \\ \hline
Buffers(Bus-2) & 8T & Repeaters & 12T \\ \hline
 & & 4-2 Translators & 16T \\ \hline
 & & 2-4 Translators & 12T \\ \hline
\end{tabular}
\end{center}
\end{table*}

\subsection{Experimental Method}
Our experimental method consists of transistor level (SPICE) simulation of the
candidate (fig.~\ref{fig:mvl_switchbox}) and the base architecture
(fig.~\ref{fig:base_switchbox}), by varying the track length, and comparison of
energy and delay in both cases.

To ensure fairness of comparison we do the following:
\begin{itemize}
\item
The unit track length equals the width of a basic CLB+Switchbox tile of a
binary FPGA, where a CLB is 4 LUTs+FFs, and no of tracks W=64. This amounts to
approx. 46 microns. We did this estimate based on a previous design flow for
binary FPGAs~\cite{fpl07}. The same length of track is used in both cases. The
tracks are routed in metal layer 3, with minimum width and minimum spacing in
28nm FDSOI technology from ST Microelectronics.  The track capacitance and
resistance are extracted with the Cadence tool QRC~\cite{}.

\item
The delay of both binary and quaternary signals depend on the slew rate. To
ensure fairness we have used the same input slew rate in both cases.

\item
The routing buffers in binary routing architecture is assumed to be a tapered
buffer with an input stage of width 1x, and an output stage of width 4x. The
quaternary buffers are optimized but the width of all transistors are limited
to 4x during optimization.
\item
for both cases we assumed an input switch box flexibility $fc_{in}=0.25$, that
is each input connects to 16 tracks in our case. Thus in both cases input
multiplexers are 16:1. For delay simulation each unit segment in the track is
loaded with four input buffers.
\item
In the quaternary case, we also count 16(No. of Inputs) 4-to-2 translators for the input pins
and 4(No. of Outputs) 2-to-4 translators, for converting binary signals coming from
CLBs. 
 
\item
For all experiments we use a test vector where all transitions (as in
table~\ref{tab:transitions}) are equally represented, for a fair comparison of
dynamic and leakage energy.

\end{itemize}

\section{Results}
\label{sec:results}
\subsection{Energy \& Delay}
\begin{figure}[h]
   \centering
   {
   \subfigure[Energy-Delay Plots ]{\includegraphics[width=0.5\textwidth,angle=0]{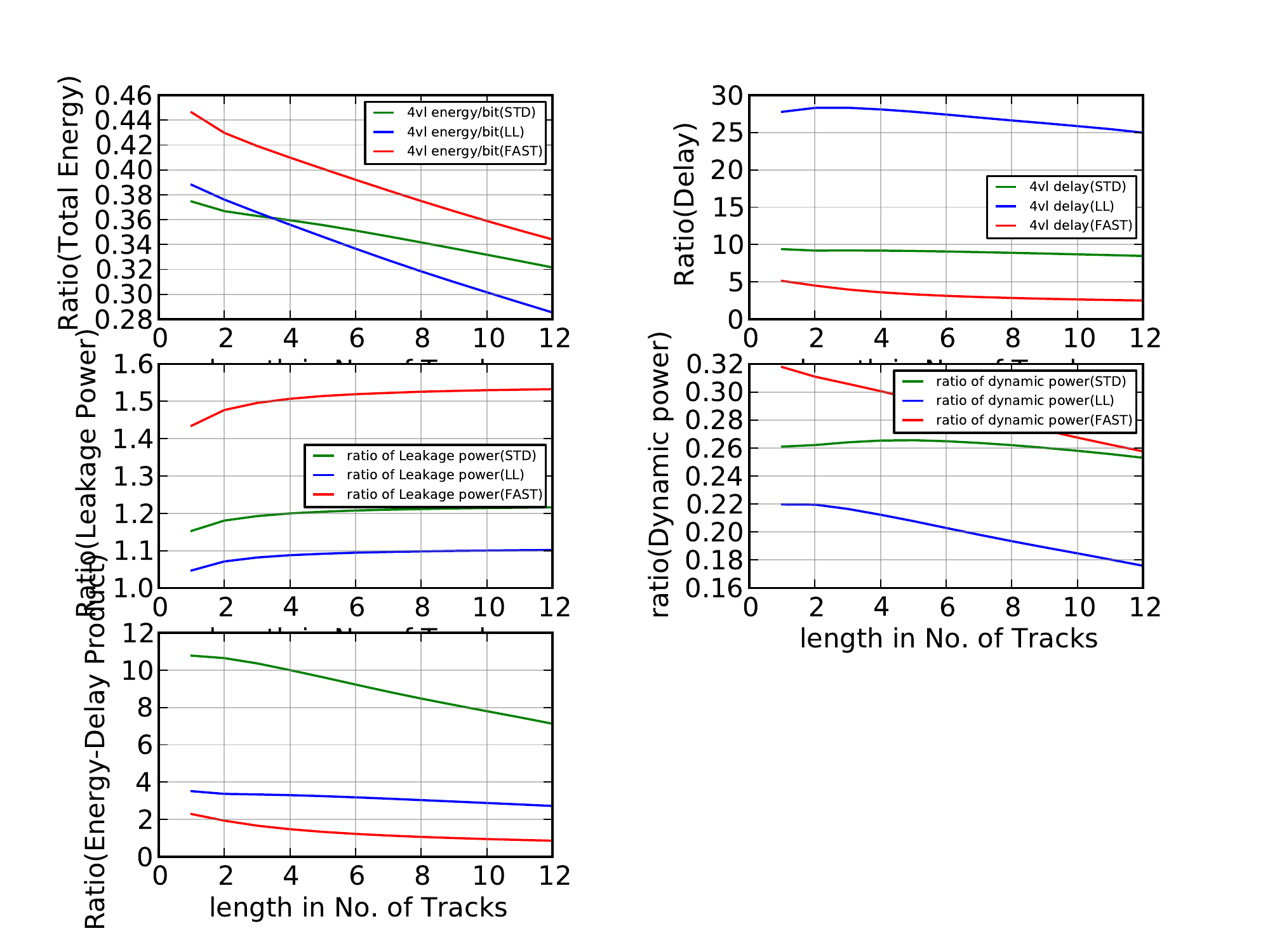}
   \label{fig:repeater2}}}
   \caption{Energy-delay plots with varying routing track length, in multiples of unit track  length. The plotted value is the ratio with
equivalent binary architectures. Cycle time used for simulation 10ns for the FAST, 40ns for the STD, and 70ns for LL (Low Leakage)}
   \label{fig:results_tracks}
\end{figure}
Figure~\ref{fig:results_tracks} plots the energy and delay comparison of the
quaternery routing tracks with that of binary tracks. We have optimized the
4-valued repeater circuits by varying the back-biasing with various different
goals, namely FAST is optimized for delay, and LL is optimized for low leakage.
STD is a compromise between the above two.

The plotted values are ratios of equivalent quantities w.r.t the value measured
in the base architecture. The delay is measured from the switchbox multiplexer
input to connection-box multiplexer output. We can see the energy and delay
ratios for FAST and LL buffers are similar to those for quaternary tracks
depicted in figure~\ref{fig:results}.  except for leakage which is less.
This is due to extra transistors in the binary bus-2 multiplexers which are
absent in quaternary case.  The leakage from these transistors compensate the
high leakage of repeaters in part. For more details about multiplexer leakage
please refer to~\cite{mux_leakage}.

The FAST buffer is optimized for delay by using higher FBB (Forward Body Bias),
consequently it has higher leakage compared to other buffers.

For the LL repeater there is very little FBB in the driving transistors which
reduces leakage, but the delay is much higher because of lower drain current
$I_D$.

We can also note that even for FAST buffer the delay is twice that of binary
signalling. The reader might note that this is not necessarily a penalty, as a
particular computation (e.g adder) synthesized in 4-valued logic will have a
smaller critical path in terms of interconnect hops. However in this article we
only concentrate on routing delays, and additional advantages of using a
complete mutli-valued arithmetic/logic are out of scope of this article.

\subsection{Transistor Count}
Based on the architecture described in section~\ref{subsec:arch} and the
transistor counts for the basic blocks (see table~\ref{tab:tcount}), we compare
transistor count for routing resources in table~\ref{tab:compare}.  The mux
count for each switchbox can be seen in figures~\ref{fig:base_switchbox}
and~\ref{fig:mvl_switchbox}. For each tile there are 16 input connection box
Muxes (16:1).
\begin{table}[h!]
\begin{center}
\caption{Routing Resources for 64 pairs of single driver tracks \& $Fc_{in}$ of 0.25. Each CLB is having 16 inputs equally distributed on all sides.
	LOF is the \emph{layout overhead factor} assumed to be $1.1$}
\label{tab:compare}
\begin{tabular}{|c|c|c|}
\hline
\multicolumn{3}{|c|}{\textbf{Binary FPGA with Bus-2 tracks}} \\ \hline 
Resource & count & Transistor Count \\ \hline 
Switchbox Muxes (BUS-2) & $4\times 64$ & $4 \times 64 \times 40T$ \\ \hline 
Buffers & $16 \times 64$  & $16 \times 64 \times 4T$  \\ \hline
Connection Box Muxes & 16 & $16 \times 76T$  \\ \hline
TOTAL &  Transistors & 15552T \\ \hline
 \multicolumn{3}{|c|}{\textbf{Quaternary FPGA}}  \\ \hline
Resource & count & Transistor Count \\ \hline
Muxes 8 input  & $4 \times 64$ & $4 \times 64 \times 26T$ \\ \hline
4-valued Repeaters & $8 \times 64$ & $8 \times 64 \times 12T \times LOF $\\ \hline
Connection Box Muxes & 16 & $16 \times 46T$ \\ \hline
4-2 Translator & 16 & $16 \times 16T \times LOF$ \\ \hline
2-4 Translator & 4 & $4 \times 12T \times LOF$ \\ \hline
TOTAL & Transistors & 14484T  \\ \hline
\end{tabular}
\end{center}
\end{table}

From the above calculation we can see that even though the 4-valued repeaters
are 1.5 times bigger in transistor count, the overall area reduction in the
routing resources are of the order of 10 \%. This is because quaternary tracks
require less routing multiplexers.
To account for layout complications we added a factor \emph{layout overhead factor}.
The layout overhead factor is around 10\% (see sec. ~\ref{sec:layout}).

\section{Variability \& Reliability}
\label{sec:var}
\subsection{Process Variation}

\begin{figure}[h]
   \centering
   \includegraphics[width=0.4\textwidth,angle=0]{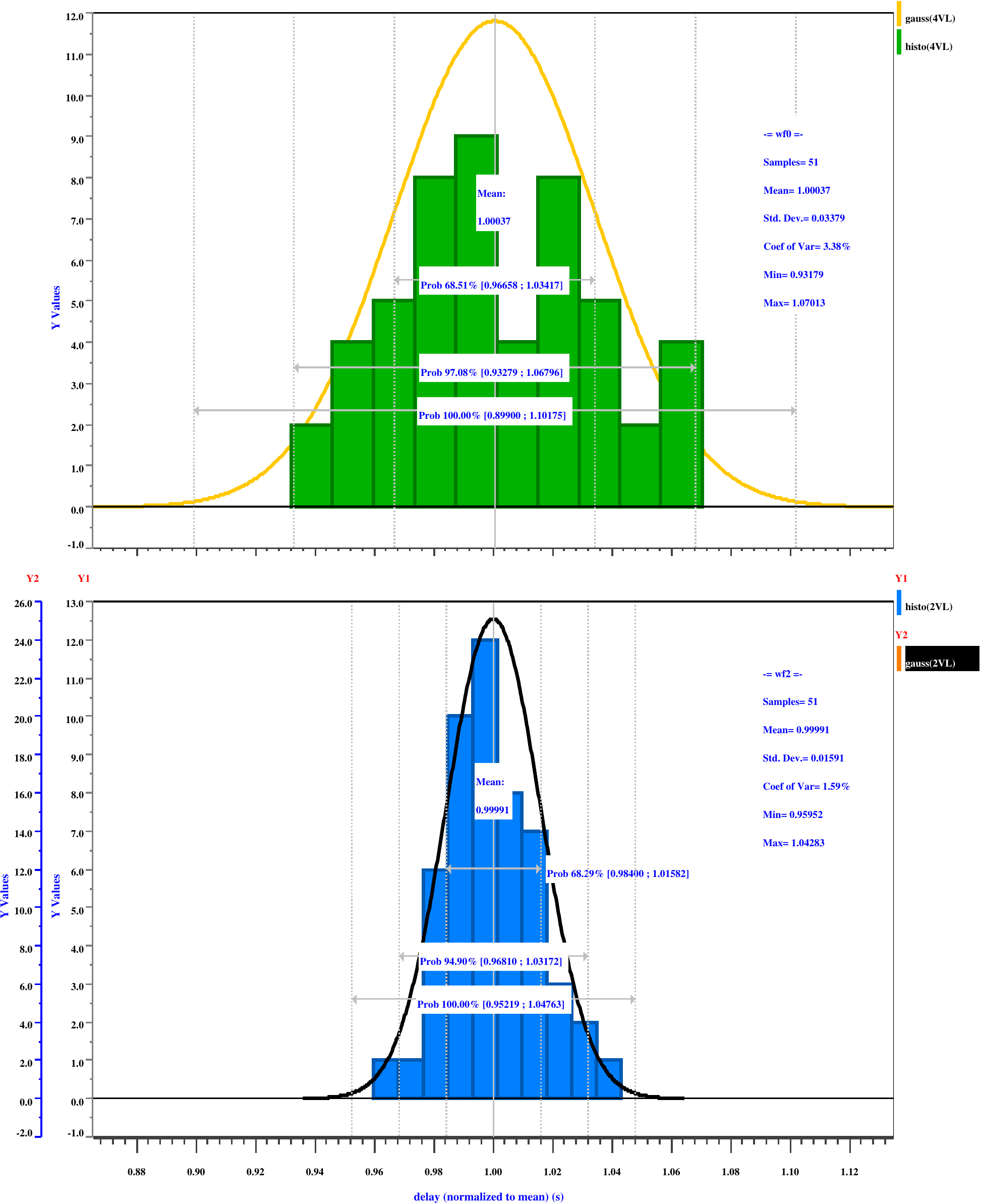}
   \caption{Results of Monte-Carlo simulations(ELDO) for quaternary (top) and binary(bottom) rouitng. X-Axis is delay and Y-axis is frequency.
   A Gaussian Distribution is fitted to the histogram generated from ELDO simulations. Quaternary Line (Length 6), $\sigma/\mu=3.38\%$, Binary Line (Length 6 ), $\sigma/\mu=1.60\%$}
   \label{fig:leakage}
\end{figure}

Process variation and signal integrity are increasingly important in
recent technology nodes. FDSOI benefits from reduced variability
thanks to absence of RDF (Random Dopant Fluctuation). However in the
quaternary repeater circuits several transistor pairs are operated
with VDD very close to the threshold Voltage, in that sense they are
working in the near-threshold region. This makes them slightly more
susceptible to process variation.

To evaluate this we conducted Monte-Carlo experiments with a length-6 tracks (hex Lines).
We see that for quaternary tracks the Pelgrom coefficient of variation ($\sigma/\mu$) is $3.38$ which is roughly twice that of the binary tracks ($1.6$).

\subsection{Sensitivity to Back-Bias Voltage}
Sensitivity to back-bias
voltage is a major concern for reliability.
We conducted a sensitivity analysis w.r.t back bias voltage $V_{BB}$ for the repeater and translator circuits.
The circuits are functional upto a 10\% variation in the $V_{BB}$. Recall that the sensitivity of $V_{th}$ to back bias volatge
is 85 mv/V.  However since very little current is drawn through the 
back bias port ($\sim$pA) thanks to insulation, there is no reason to expect significant IR DROP in 
$V_{BB}$ even in a big chip. There are other concerns such as transient noise on the back bias port but 
it is out of scope of this article.


\section{Layout}
\label{sec:layout}
\begin{figure}
\centering
\includegraphics[width=0.4\textwidth]{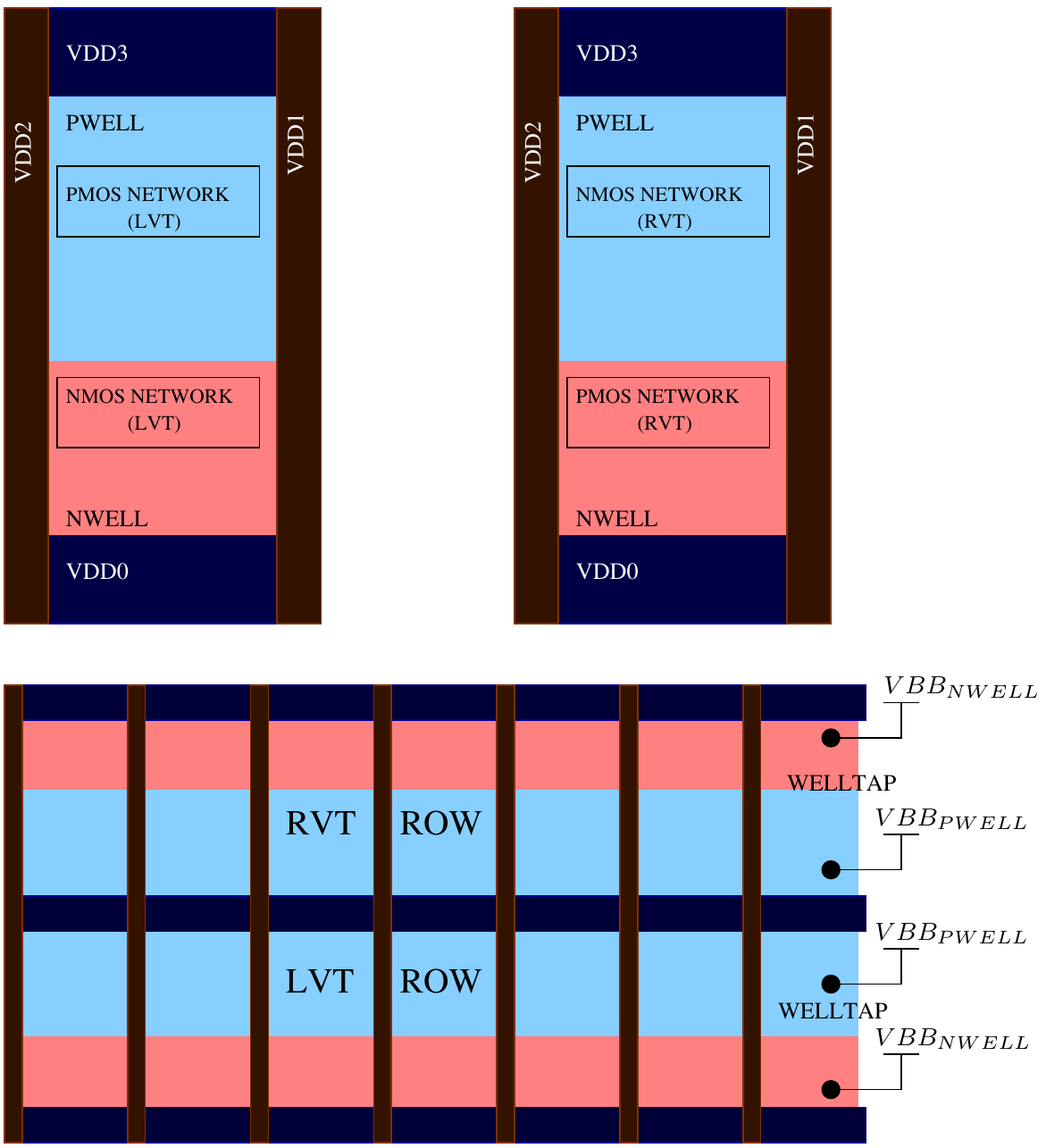}
\caption{Transistor placement on the layout, Deep blue shows metal level 1, and brown is metal level 2. Back biasing is done via 
a welltap cell.}
\label{fig:layout}
\end{figure}

In fig.~\ref{fig:layout} we present an outline of the transistor/std. cell
placements to implement multi-valued circuits. As seen in the circuit diagram(fig. ~\ref{fig:repeater4})
we have used both RVT (for Reverse Body-Bias) and LVT (for Forward Body Bias) 
transistors. Mixing LVT and RVT transistors generates a complicated layout 
as observed in ~\cite{prayer}. However in our case we have an advantage that
the back-biasing voltages are fixed, and we use the same $V_{BB}$ for LVT and RVT 
transistors in PWELL and NWELL. Thus we can arrange RVT and LVT transistors
in back to back rows which share a well. In this scheme we don't need and 
extra isolation between these rows. A similar layout scheme is prescribed in ~\cite{prayer}.

Also we can see that the VDD2 and VDD1 supply lines are laid out vertically in metal level 2, and 
VDD3 (VDD in standard designs)  and VDD0( GND in standard designs) are laid out in 
horizontal direction in metal level 1 as in  traditional STD. Cell flows.  The amount of routing area taken by the 
extra supply lines $\{VDD2,VDD1\}$ does not exceed 15\% of the cell routing area in metal level 2.
Given 6 signal routing layers we can expect a gain of $(5 \times 0.5+1 \times 0.35=0.47)$ 47\%  gain in overall wire routing area.

For design rule concerns the actual layout of the repeater is roughly 10\% bigger than the layout 
size calculated based on transistor counts, so we use a layout factor (LOF) of $1.1$ in table~\ref{tab:compare}
to calculate the gain in transistor area.

\section{Conclusion}
\label{sec:conclusion}
Based on the above experimental results, for Quaternary signalling in an FPGA
we can expect
\begin{itemize}

\item
40-50\% reduction in wire routing area.
\item
$\sim$10\% reduction in transistor  area occupied by routing resources which is a major share
of FPGA area.
\item
3x reduction in dynamic power consumption, which is equivalent to CMOS
operation in near-threshold~\cite{near_vth}.
\end{itemize}

There is also a penalty of leakage power. This can be mitigated by the use of
low-leakage versions, or use of sleep modes.

Because of the above characteristics we can imagine the following usage
scenarios for multi-valued tracks
\begin{itemize}
\item
Bus based FPGAs~\cite{rose_bus} have been reported where a bus of 4 gives the
optimum area efficiency.  These bus based tracks can be implemented as
multi-valued tracks.
\item
Higher order routing topologies (e.g Butterfly Fat Tree, Hypercube) which are
limited by wire routing area~\cite{bft_dehon}.
\item
Asynchronous logic which requires dual rail signalling and thus demanding in
terms of routing~\cite{furber_ternary}.
\item
Ultra low-energy FPGAs.
\end{itemize}

There could also be some concerns regarding the susceptibility to process
variation. Several methods at higher levels of CAD are
proposed~\cite{leihe,dehon} to mitigate these problems. These methods have to
be used in conjunction with multi-valued tracks. 


Our future research direction will be to do a benchmark based study for FPGA
architectures using multi-valued tracks, and leakage optimization of the
circuits presented in this article.

\bibliographystyle{abbrv}
\bibliography{mvl}

\newpage
\onecolumn
\section*{Anonymous Reviews}
\subsection*{Reviewer 1}
\begin{verbatim}
I thought the idea in this paper was interesting but there are a large number
of issues that need to be corrected.

Counting area in transistors is not an acceptable way of comparing area between
the baseline and new architecture. As shown in Fig 1, the FDSOI with body bias
requires a body contact for each unique set of transistors sharing a Vdd. This
can potentially increase the area per transistor by perhaps 50\% in the case of
individual transistors. The paper needs a better area model that includes this.

I am not sure the paper deals fully with Vt and other variation. In addition to
the random variation of zero bias Vt0, there is also the variation of the body
effect between transistors. Since the paper is using body bias to adjust Vt by
~600mV, even a 10\% variation in body effect could add another 60mV of Vt
variation to each transistor beyond the variation at zero bias. Further, using
multiple power supplies creates the possibility of double ended variation. For
example with 50mV Vdd variation the 900mV supply could droop to 850mV, the
600mV supply boost to 650, leaving only 200mV Vgs for the Vdd to 2/3Vdd modes.
After adding in the two sources of Vt variation to the power supply variation,
are the circuits even operational?

I'm not sure why the routing muxes are shown as being controlled by quaterny or
standard Vdd binary. Since routing is controlled statically, the simplest
approach is to use a RAM cell powered by the Vdd boost voltage.

The paper makes several analyses of delay using power law models. I understand
that these are related to velocity saturated transistor drive currents and do
not think they are likely to be accurate in a low-Vdd regime. The only results
I would trust are SPICE models. SPICE is not mentioned in the paper anywhere so
it is not clear to me how the results were generated. Similarly it is unclear
how the leakage results are generated.

The paper fails to make an complete comparison between binary and quaternary by
only comparing the two power-delay points. Binary is much faster but 3X higher
power. However if I drop the Vdd of binary by a factor of 0.6X, it will slow
down and the dynamic power will reduce by 3X (0.6^2 = 0.36, about 3X
reduction). I will hazard a guess that binary at 0.6X Vdd is faster and same
power as quaternery, but this needs to be evaluated.
\end{verbatim}

\subsection*{Reviewer 2}
\begin{verbatim}
multiple values on sinals wires have been discussed for FPGAs for many years.
Most papers exploite the bandwidth by time-multiplexing rather than
multi-values.  Some discussion of the advantages of your method is warranted,
particularly since the tighter control required for multi-values is difficult
in advanced nodess.  Also your assumed performance permits significant time
multiplexing of binary values.

Section II.  A description of FDSOI is only necessary if your subsequent
analysis depends on it.

Where does table 1 come from?  The units in table 2 are unclear.
Fig 3 bubbles on inverters are much too small.

I have serious problems with your experimental procedure.  You propose a binary
architecture with the same architectural limitations imposed by your
multi-valued signaling.  Basically, you wipe out your major drawaback: the fact
that multi-valued signals must be routed together to share a wiring trace.
Modern binary FPGAs do not do that and achieve a performance, power and density
advantage.

This is absolutely not "fairness" (V.B paragraph 2).  Other architectures don't
need to do it.        The penalties (performance, power, size) for bundling and
un-bundling signals needs to be incorporated into your result, but not the
result for the binary alternatives.

With this in mind, you need to report performance differences.        Ideally,
these would include those imposed by the architecutre, after place and route.
\end{verbatim}

\subsection*{Reviewer 3}
\begin{verbatim}
The authors propose using body-biased FDSOI technology to create multi-valued
routing tracks for FPGAs. The body bias is used to create transistors with
different Vt values that can respond to different voltage levels (they do
3-valued logic with 1/3 of Vdd steps, and 4-valued logic with 1/4 Vdd steps).
The logic on the FPGA is still binary: only routing (or a subset of it) is
proposed to be multi-valued. They claim that the routing area would decrease
both due to a reduction in the number of wires, as well as area reduction of
the buffers and repeaters. The circuits are explained well and the delay and
power models presented in Sec IV make sense.

The trouble starts with Section V. I expected to see SPICE simulations to
validate the idea. Vth values are bound to have variations. The graphs of Fig 8
do not tell the whole story. Is a chain of repeaters going to switch correctly?
Monte-Carlo simulations of the chain are needed.

Experimental results showing noise susceptibility have to be included. The
graphs of Fig 9 are inadequate: the authors could have done Monte-Carlo SPICE
simulations, modeling corss-talk and supply noise.

They only mention (in passing) methods to tackle cross talk, such as spacing
apart the wire segments more or limit the speed of operation. That would result
in partly defeating the purpose of multi-valued logic that they used to justify
their method in the first place. The effects of spacing and lower frequency to
achieve similar noise immunity as binary logic should be quantified and
compared.

Overall, I liked the idea presented in the paper, but I feel it needs to be
improved more to warrant publication. Even if the paper turns out to be a
negative-results paper, I d like to see it fleshed out.  
\end{verbatim}

\subsection*{Reviewer 4}
\begin{verbatim}
The main premise of using multiple voltages to reduce the wire count is
interesting. In general, a technique like this would make more sense in a board
where high quality wires are a rare, valuable resource. But applying the
technique within a chip isn’t unheard of. And the truth is that wires are not
scaling like transistors. Even if the metal design rules come down with
technology, scaling wires with the design rules kills performance. Modern
device costs can be dominated by metal, and more so for programmable devices
than ASICs. So reducing wires is important. That said, my biggest concern for
this paper is just the cost overhead. I expect the cost (including area, power
and delays) to driving and receiving these multi-voltage signals to be
enormously prohibitive compared to simple buffers. For me, the other issues are
secondary, and I’d like to see the costs addressed up front and convincingly.
\end{verbatim}

\end{document}